\documentclass[letterpaper, 10 pt, conference]{ieeeconf}  % Comment this line out if you need a4paper

\IEEEoverridecommandlockouts                              % This command is only needed if 
\title{\LARGE \bf
Anticipation in Human-Robot Cooperation: 
A Recurrent Neural Network Approach for Multiple Action Sequences Prediction
}

\author{Paul Schydlo$^{1}$, Mirko Rakovic$^{1, 2}$, Lorenzo Jamone$^{3}$ and Jos\'e Santos-Victor$^{1}$% <-this % stops a space
\thanks{$^{1}$Institute for Systems and Robotics, Instituto Superior T\'ecnico, University Lisbon, Portugal}%
\thanks{$^{2}$Faculty of Technical Sciences, University of Novi
Sad, Novi Sad, Serbia}%
\thanks{$^{3}$Queen Mary University, London, United Kingdom}}%

\usepackage{cite}
\usepackage{graphicx}
\usepackage{amsmath,amssymb}
\usepackage{subfig} 

\graphicspath{{images/}}

\usepackage[printonlyused]{acronym}

\newacro{RNN}{Recurrent Neural Network}
\newacro{LSTM}{Long Short Term Memory}

\begin{document}

\maketitle
\thispagestyle{empty}
\pagestyle{empty}

%%%%%%%%%%%%%%%%%%%%%%%%%%%%%%%%%%%%%%%%%%%%%%%%%%%%%%%%%%%%%%%%%%%%%%%%%%%%%%%%
\begin{abstract}
Close human-robot cooperation is a key enabler for new developments in advanced manufacturing and assistive applications. Close cooperation require robots that can predict human actions and intent, understanding human non-verbal cues. Recent approaches based on neural networks have led to encouraging results in the human action prediction problem both in continuous and discrete spaces. Our approach extends the research in this direction.

Our contributions are three-fold. First, we validate the use of gaze and body pose cues as a means of predicting human action through a feature selection method. Next, we address two shortcomings of existing literature: predicting multiple and variable-length action sequences. This is achieved by applying an encoder-decoder recurrent neural network topology in the discrete action prediction problem. 

% by comparing the relative accuracy of a model trained on a body pose only and a combined pose and gaze feature set

In addition, we theoretically demonstrate the importance of predicting multiple action sequences as a means of estimating the stochastic reward in a human robot cooperation scenario. 

Finally, we show the ability to effectively train the prediction model on an action prediction dataset, involving human motion data, and explore the influence of the model's parameters on its performance.

%Close human-robot cooperation is an important research area in an age where industrial automation is on the rise. Close cooperation requires the understanding of human non-verbal cues and predicting human actions and intent.

%1Neural network based approaches have had success in the human action prediction problem both in continuous and discrete space.

%This paper expands on the existing literature addressing two shortcomings, variable prediction length and multiple trajectory prediction, by introducing the encoder-decoder architecture to the discrete action prediction problem.

%We theoretically demonstrate the importance of predicting multiple trajectories as a means of estimating the stochastic reward in a human robot cooperation scenario.

%Furthermore, we are able to effectively train the encoder-decoder model on an action prediction dataset and exploring the influence of the beam-width and the context vector dimensionality on the model's performance.

\end{abstract}

%%%%%%%%%%%%%%%%%%%%%%%%%%%%%%%%%%%%%%%%%%%%%%%%%%%%%%%%%%%%%%%%%%%%%%%%%%%%%%%%
\section{INTRODUCTION AND RELATED WORK}

In a world with a growing number of autonomous systems and moving towards the coexistence and cooperation between humans and sophisticated robots, it is crucial to enable artificial systems to understand and predict human behaviour. This ability finds applications in areas such as cooperative robotics \cite{Huang2016e, Sakita2004a}, auto-mobile safety \cite{Jain2016}, elderly care \cite{Yonezawa2013}, among many others \cite{Admoni2016}.

In addition to the use of speech for communicating and coordinating their next actions, humans rely extensively on  non-verbal cues for action and movement prediction~\cite{Huang2015a}. Situations where fast cooperation is essential, for example cooperative assembly, require the understanding of subtle non-verbal cues \cite{Sakita2004a} about the human intention and future action. In these scenarios, it is not enough to merely recognize the current action. Instead, it is fundamental to  predict actions and anticipate the intent in order to guarantee seamless cooperation \cite{Sebanz2009}.

\subsection{Non-verbal cues}

There are several non-verbal cues that enable human action prediction \cite{Breazeal2005a, Argyle1973}. This paper takes into account two of them: gaze and body posture. Gaze is important, as it has both a role in social communication in conveying turn taking behaviour \cite{Ho2015} or attention in conversation, but at the same time it is deeply related to the agent's Theory of Mind \cite{Baron-Cohen1997} about the collaboration partner and codifies the action goals through both visuo-motor coupling \cite{Lukic2014b} and attention \cite{Argyle1973}. Body posture, similarly to gaze, can serve both a social and intention conveying signal while also indicating possible action targets.

Past works have focused on either gaze \cite{Jain2016, Sakita2004a, Huang2016e} or body pose \cite{Perez-DArpino2015b} cues and their relation to action recognition and prediction. Both are important in understanding human behaviour and give information about the human's action goal. % Can their importance be quantified?

Research on non-verbal cues in human-robot cooperation has a long history, including the bulk of work on mirror neurons \cite{GALLESE1998493} and its computational and robotic models and implementations \cite{lopes2005visual}. Relevant work include Admoni \cite{Admoni2016} use of human gaze as a means of estimating the human intent, modelling the relation between the gaze and the action goal by their relative distance. Huang \cite{Huang2016e} quantified the importance of gaze features, successfully demonstrating the importance of gaze by proactively planning actions according to the human intent. 

\subsection{Prediction models}

Human action prediction can be solved at different levels of abstraction and is concerned with estimating a probability distribution over the set of next possible actions. 

At a higher level of abstraction, models can predict actions in a discrete space\cite{Jain, Jain2016} where the actions are symbolic in nature and can represent underlying movement patterns, e.g. ``press-button'' or ``grab-object''. On a lower level of abstraction, movement can be directly anticipated in a continuous space \cite{Martinez2017}, e.g. human walking trajectories. 

Predicting in continuous space has been addressed in the context of body pose and human trajectory prediction. Relevant work include the use of Recurrent Neural Networks by Martinez \cite{Martinez2017} as a means of predicting coherent future joint trajectories.  % text about continuous action prediction

%It is of greater interest to obtain the distribution instead of just a possible outcome in this space.

The dual problem is action prediction in discrete outcome space. Relevant work include a Conditional Random Field based approach by Koppula \cite{Koppula2016} to capture temporal dependencies and Saponaro's Hidden Markov Model based approach \cite{Saponaro2013a}. Recently, Recurrent Neural  Networks, without  limiting  Markovian  assumptions, have shown excellent results [16], [17], [20]. Relevant work include, the structural RNN as a means of encoding past contextual information and predicting a fixed number of steps in the future by Jain \cite{Jain}. While the field has had a rapid evolution in the last couple of years, there are two shortcomings in the literature this paper addresses.
%Alahi \cite{Alahi2016} applied a \ac{LSTM} structure to study the effect of coupling agent's trajectory predictions.

The first is concerned with predicting a fixed versus a variable number of steps into the future. While models like\cite{Jain} have a remarkable ability to condense contextual past information, their scope is limited to fixed step ahead prediction length. This paper aims at extending discriminative recurrent models in a classification setting with variable length action sequence prediction.

The second shortcoming is related to the single future action sequence versus multiple future action sequences. While models like the one introduced in \cite{Martinez2017} are able to effectively use recurrent models to predict a variable number of steps into the future, their scope is limited to a regression setting, where sampling multiple future action sequences is a non-trivial problem. This paper explores a multiple future action sequence predictor in the classification setting.

%Expanding the future trajectory space of discrete action outcomes has an exponential complexity. This paper introduces the beam search decoder as a way to prune the exponentially growing future action outcome space.
 
\subsection{Contributions}

The main contributions of this paper are the following:
\begin{itemize}  
	\item Quantifying the \textbf{relative importance of pose and gaze} features in an intention recognition scenario.
	\item Extending recurrent neural network fixed step action prediction with 	\textbf{variable length action prediction}.
	\item Introducing the simultaneous prediction of \textbf{multiple future action sequences}.
\end{itemize}

%\section{RELATED WORK}

%This section is split in two parts. The first is concerned with giving an overview of the literature related to the use of non-verbal cues in anticipating human action. The second section reviews prediction models applied to the problem of action anticipation.

%\subsection{Non-verbal cues}

%[Ainda faltam citations]\\

%\subsection{Anticipation Models}

%As previously mentioned, an important distinction are anticipation models in discrete and continuous outcome space.

%These can then be further divided as methods which look at the semantics of the scene and those which look at the human movement patterns.

%[Ainda faltam citations]\\

%The focus of this paper is anticipation in discrete outcome space and methods which do not take into account scene semantics such as interaction affordances and places of interest

\section{APPROACH}

Our work looks at the action prediction problem from an end-to-end perspective, starting with the problem of non-verbal cues selection and moving on to develop an action sequence prediction model. Keeping in mind the final goal, predicting future human action given a sequence of past non-verbal cues such as gaze and pose, this section is organized in a sequential bottom-up order. 

First, we address the issue of establishing a quantitative metric for assessing the relative importance of pose and gaze features. Then, in Section II.B, we introduce the multiple action sequence prediction model which is one of the key contributions of this paper. Predicting action sequences introduces complexity issues which are handled in Section~II.C.
Finally, in Section III we use the distribution over future action sequences sampled from the model, introduced in Section II, to estimate the expected future reward in a human-robot cooperation scenario. 

%This paper starts with introducing the human action prediction problem from a feature selection perspective moving on to laying out the architecture of the encoder-decoder discrete action prediction model and finishes with the modelling of a Dec-MDP cooperation scenario which applies the action prediction model introduced in the paper.

\subsection{Feature importance}

This section seeks to introduce a quantitative metric for the relative gaze and body pose cues importance, two commonly used features in non-verbal communication \cite{Breazeal2005a}. Selecting the right features is an important step to reduce the complexity and increase the robustness of our models. 

There are different feature selection methods which can be categorized into \emph{filter}, \emph{wrapper} and \emph{embedded} classes~\cite{Saeys2007}. Since the relation between the features is unknown, it is assumed to be non-linear in nature. Following the non-linearity assumption, the focus of this section will be on the \emph{wrapper} class of feature selection methods. This class of methods captures non-linear relation between the variables through a black-box model. It starts by training the model on subsets of the feature space and then ranks the features according to the model's accuracy\cite{Saeys2007}.

In the case of this paper, the black-box model is the intention recognition model, a \ac{RNN} sequence to sequence model. The structure of the model is defined by an embedding layer, which at every step transforms the feature vector into an intermediary representation, acting as an input to the model's \ac{RNN}. For every input, this \ac{RNN} returns a discrete distribution over intentions. This distribution is obtained by projecting the recurrent neural network's internal state and normalizing it through a softmax layer.
\begin{figure}[htbp]
	\centering

	\includegraphics[scale=0.035]{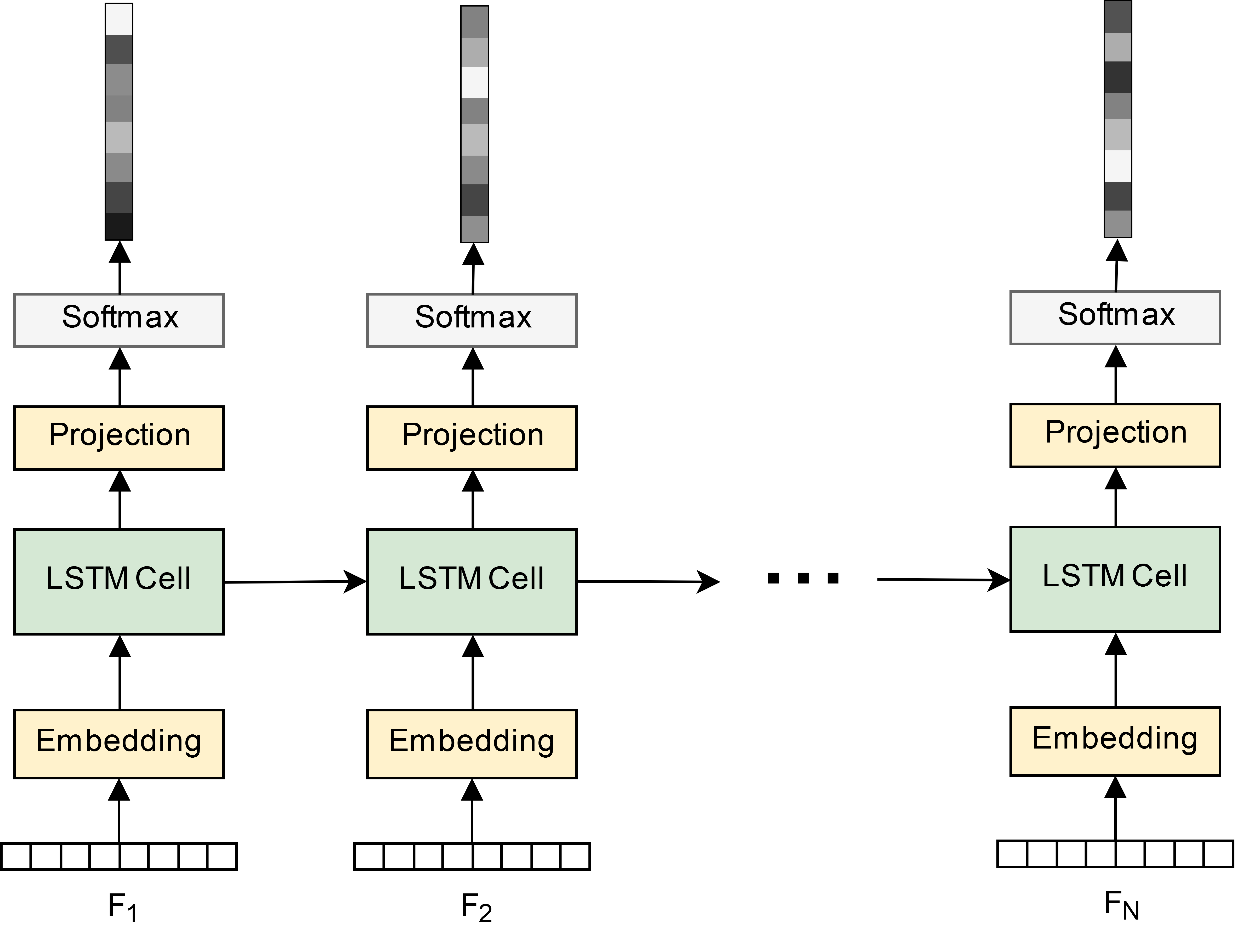}
	\caption{\textbf{Intention recognition model.} This model maps a sequence of input features to a sequence of discrete distributions over the action vocabulary.}
	\label{recognitionrnn}
\end{figure}

The prediction accuracy of the model with and without a given feature can be considered a proxy for the feature's added information. Having established a quantitative measure of the gaze and pose features' importance, the next section introduces the prediction model.

\subsection{Prediction model}

This section introduces the discrete encoder-decoder recurrent neural network topology which seeks to solve the shortcomings enumerated in section II. The first part of the model is a contextual information encoder. The encoder condenses past information into a fixed length context vector through a \ac{LSTM} cell. The embedding is a fully connected layer (FeatureVectorDim~$\times$~50), where FeatureVectorDim is the size of the feature vector. The embedding layer includes dropouts which act as a regularization to the model \cite{Srivastava2014}. The encoder \ac{LSTM}'s hidden state dimension is 20. This context vector, the internal state of the encoding \ac{LSTM}, is the initial state of the second part of the model, the decoder. 

The decoder is responsible for generating a coherent future sequence of actions. At each step the decoder, an \ac{LSTM} cell, returns a discrete distribution over possible future actions. This distribution is obtained by projecting the decoder's internal state and normalizing it using a softmax layer. The decoding process samples an action from the distribution and feeds it back as an input to the next decoding iteration. The projection is a fully connected layer (HiddenStateDim x VocabDim), where HiddenStateDim is the size of the hidden state, 20, and VocabDim the dimension of the action discrete possible actions vocabulary, 11. The decoder \ac{LSTM}'s hidden state dimension is 20.
\begin{figure}[htbp]
	\centering
	
	\includegraphics[scale=0.3]{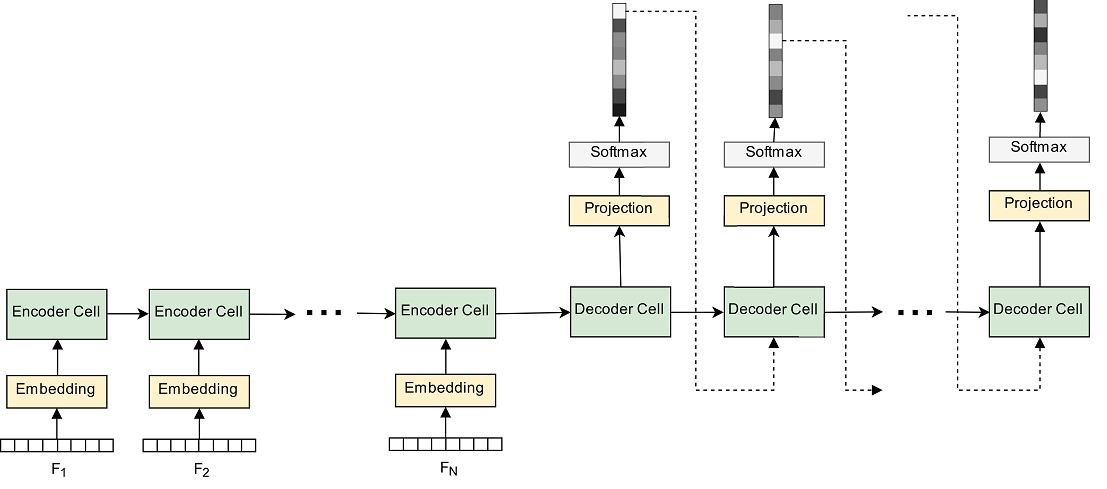}
	\caption{\textbf{Encoder-decoder model.} The left part summarises past information into a fixed length context vector. Right part expands this context vector into future action sequences.}
	\label{beamdecoder}
	\vspace{-1em}
\end{figure}

The model is trained with the Adam algorithm using a sequential cross entropy loss. The cross entropy cost (\ref{eq:cross_entropy}) is a measure of difference between two distributions: predicted distribution, p, and reference distribution, r. The discrete distribution is defined over the limited set of possible actions, A, where every possible action, a, is an instance of this set,  p(a) and r(a) define respectively the predicted and reference probability of the action, a. The sequential cross entropy is obtained by summing the cross entropy, H, cost over the prediction steps:
\begin{equation} \label{eq:cross_entropy}
H(p, r) = -\sum_{a \in A} p(a) \log \left( r(a) \right).
\end{equation}

%The sequential cross entropy is obtained by summing the cross entropy cost over the prediction steps. Here P and Q represent respectively the predicted and reference sequence of discrete distributions and N the prediction length. 

%\begin{equation} \label{eq:seq_loss}
%	Loss(P, Q) = \sum_i^N H(P_i, Q_i)
%\end{equation}

After training, the decoding process allows for variable length action sequence prediction. Expanding every possible future action sequences is NP hard and computationally intractable. The next section looks more closely at this issue and introduces one possible solution to the problem.

\subsection{Complexity issues}

The previous section hints at the complexity underlying the decoding process. At every decoding step, the decoder samples one or more actions from the output distribution as possible actions at a given time step; it then expands these actions by branching and feeding them individually as input to the next decoder iteration. There are two strategies that could be applied to this decoding process. 

Naively expanding the space of all possible action sequences and selecting the most probable action sequence at the end seems like a reasonable idea. Nevertheless, expanding the actions at each step results in a vocabulary sized multiplier in the number of possible action sequences at every prediction step. In terms of complexity this means that the number of action sequences increases exponentially with the number of prediction steps. Considering a 10 actions vocabulary size, the first decoding step results in 10 action sequences, expanding the 10 action sequences results in 100 possible action sequences for a two step ahead prediction, a N step ahead prediction would result in $N^{10}$ possible action sequences.

Greedily expanding only the best option, could be a solution to the exponentially expanding trajectory space, nevertheless it has the shortcoming that this method only returns one action sequence prediction. 

A common solution to these two problems is the implementation of a \emph{beam search} based decoder \cite{Cho2014}. This method keeps a set of the top K best future action sequences at every decoding step, expanding by the action vocabulary size and pruning the action sequence set back to the top K future action sequences. The result is a sample of the top K most probable future action sequences ordered by likelihood. These trajectories are called beams and K is the beam width parameter.
%
% \begin{figure}[htbp]
% 	\centering
% 	\begin{subfigure}{.2\textwidth}
% 		\centering
% 		\includegraphics[width=.7\linewidth]{Naive_highres}
% 		\caption{Exhaustive search}
% 		\label{fig:sub1}
% 	\end{subfigure}%
% 	\begin{subfigure}{.2\textwidth}
% 		\centering
% 		\includegraphics[width=.7\linewidth]{Greedy_highres}
% 		\caption{Greedy search}
% 		\label{fig:sub2}
% 	\end{subfigure}
% 	\begin{subfigure}{.2\textwidth}
% 		\centering
% 		\includegraphics[width=.7\linewidth]{Beam_highres}
% 		\caption{Beam search}
% 		\label{fig:sub2}
% 	\end{subfigure}
% 	\caption{\textbf{Search methods comparison.} Exhaustive search expands all possible trajectories. Greedy search picks the most probable action at every step. Beam search keeps a set of the best K trajectories, expanding and pruning the set at every step.}
% 	\label{fig:test}
% \end{figure}
%
\begin{figure}
\centering
\subfloat[][]
{\includegraphics[width=0.25\linewidth]{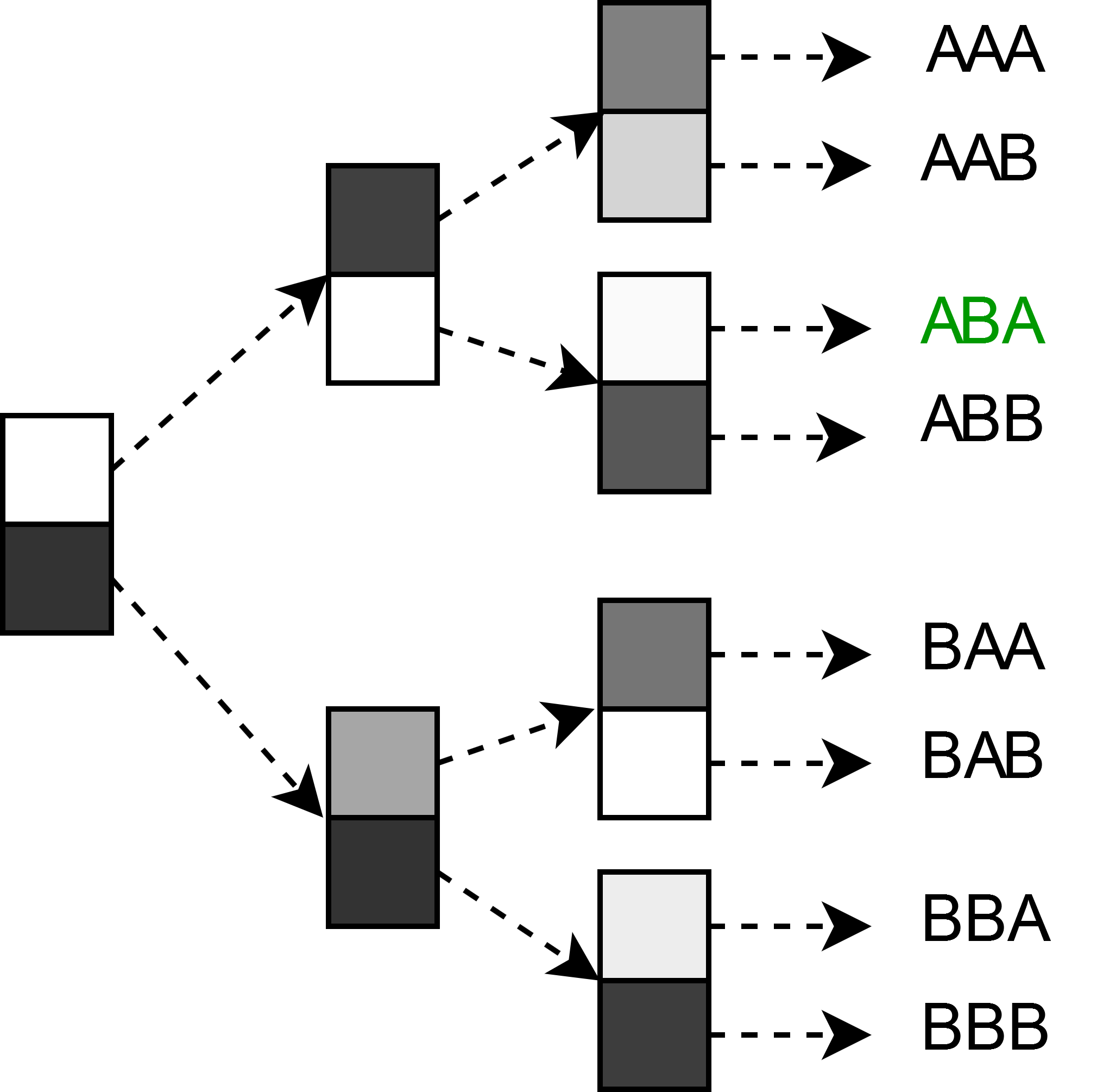}
\label{fig:first_subfig} } \quad
\subfloat[][]
{\includegraphics[width=0.25\linewidth]{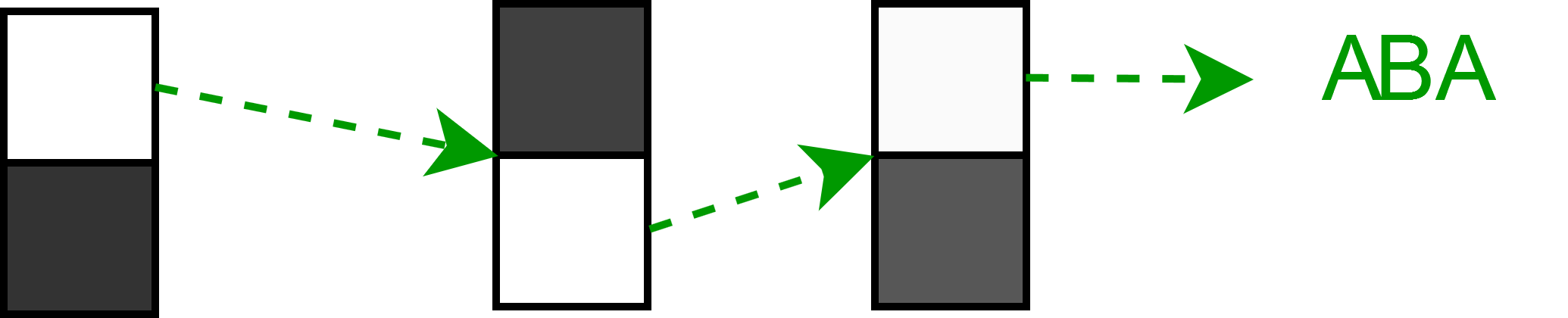}
\label{fig:second_subfig} }\quad
\subfloat[][]
{\includegraphics[width=0.25\linewidth]{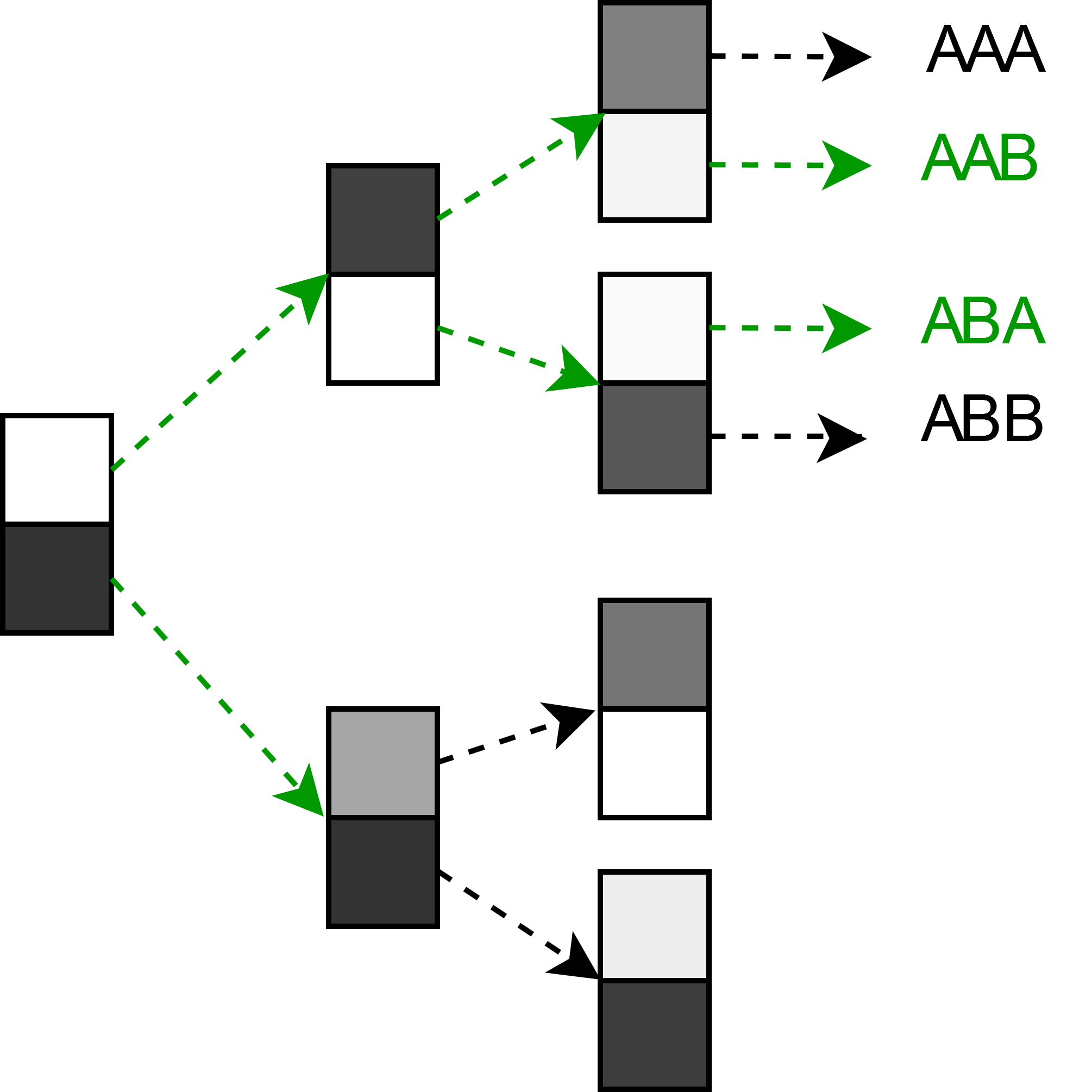}
\label{fig:third_subfig} }
\caption{\textbf{Search methods comparison.} a) Exhaustive search expands all possible action sequences. b) Greedy search picks the most probable action at every step. c) Beam search keeps a set of the best K action sequences, expanding and pruning the set at every step.}
\label{fig:search}
\vspace{-1em}
\end{figure}

\section{APPLICATION SCENARIO}

Anticipating a set of possible future actions is important in cooperative assembly scenarios, where two agents work together in a fast paced joint action setting. This scenario aims to clarify the importance and some caveats of the action prediction problem in human robot cooperation scenarios.

%\begin{figure}[htbp]
%	\centering
%	\includegraphics[scale=0.6]{iCub}
%	\caption{\textbf{Joint action scenario.} The human and robot act independently %towards maximizing a joint reward function.}
%	\label{iCub}
%\end{figure}

This setting 
%{(Fig. \ref{iCub}) 
is defined by a set of possible world states, $S$, human and robot action pairs,$A$:$(a_H, a_R)$, transition between states as a function of the current state and joint action pair, $T(S, A)$, and a joint immediate reward function, $R(S, A)$. For the sake of example, the world state could be a set of pre-conditions, T a set of action-effect axioms and R a reward function on the sub-goal completion. %[]

% Both the human and the robot have full knowledge of the final objective. 

Given an initial state, $S_0$, and an action sequence, $\textbf{A}$, i.e. a series of action pairs $(a^H, a^R)$ at N equidistant time steps, the total reward, $R_t$, is given by (\ref{eq:reward}), where $A_i$ and $S_i$ correspond respectively to the human-robot action pair and world state at time step i and, N the number of time steps:
\begin{equation} \label{eq:reward}
R_t(S_0, \textbf{A}) = \sum_{i=0}^{N} R(S_i, \textbf{A}_i).
\end{equation}

In this setting, the robot selects an action sequence, $A^R$, maximising the joint reward, $R$, and the human action sequence, $A^H$, is unknown and non-deterministic from the perspective of the robot. Therefore, the future reward associated to a chosen robot action sequence, $A^R$ can be estimated as an expectation over the set of possible human actions, $A^H$, given by (\ref{eq:expected_reward}), where  $p(A^{H,k})$ represents the probability of a human action sequence, $A^{H,K}$, $R(S_i, (a^H, a^R))$, the reward associated to the human-robot action pair in the world state $S_i$, $\#H$ the cardinality of the set of possible human actions and N the number of time steps into the future:
\begin{equation} 
\label{eq:expected_reward}
\mathbb{E}\left[R_t(S_0, A^R)\right] = \sum_{k=0}^{\#H}\left[\sum_{i=0}^{N}
R\left(S_i,\left(A^{H, k}_i,A^R_i\right)\right)\right]p(A^{H, k}).
\end{equation}

Computing the expectation, requires expanding all possible action sequences, which is computationally intractable, i.e. NP hard. We will now see how the beam search, introduced earlier, enables the estimation of this reward.

Considering the set of most probable action sequences as representative of the future human behaviour, that is, the distribution has finite variance, we can approximate the expected reward through a biased Monte Carlo estimation. This is achieved by summing and weighting the reward of a given human-robot action sequence by the human action sequence probability (\ref{eq:expected_reward_beam}). Increasing the number of predicted human action sequences, K, approximates the reward better but is computationally more demanding. Here $p(b_k)$ represents the probability of the kth beam (predicted action sequence), while  $S_i$, $b^k_i$ and $A^R_i$ represent respectively the world state, the action performed by the human in the beam k and the robot in the action sequence $A^R$ at time step i:
\begin{equation} \label{eq:expected_reward_beam}
\mathbb{E}\left[R(S_0, A^R)\right] = \frac{\sum_{k=0}^{K}\left[ \sum_{i=0}^{N} 
R(S_i, (b_i^k, A_i^R))\right]p(b^k)}
{\sum_{k=0}^K p(b^k)}.
\end{equation}

As the beam count tends to the total number of possible action sequence combinations, this expression approximates the expected reward (\ref{eq:expected_reward}).

% Here we can apply [Insert Monte Carlo method] to approximate the expected reward by importance sampling possible future outcomes.

\section{EXPERIMENTS AND RESULTS}

%In this section we review the experimental set-up used to evaluate the model. 

We start by describing the datasets used in the evaluation, we move on to compare the non-verbal cues importance and finish by evaluating the action sequence prediction model on a dataset that includes body pose information. 

\subsection{Datasets}

\begin{figure}
\centering
\subfloat[][]
{\includegraphics[width=0.355\linewidth, trim=-40 0 0 1mm]{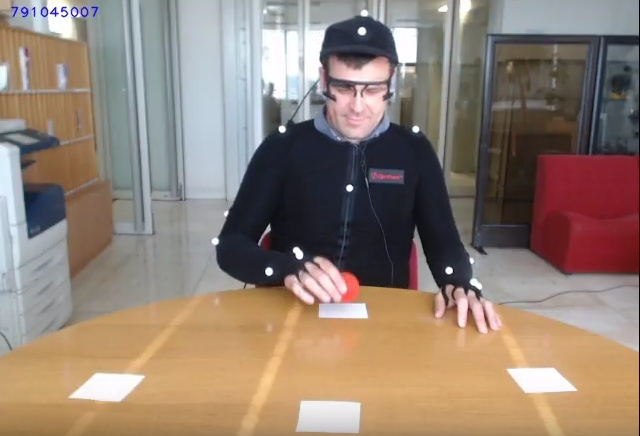}
\label{fig:acticipate} } \quad
\subfloat[][]
{\includegraphics[width=0.34\linewidth, trim=-40 0 0 1mm]{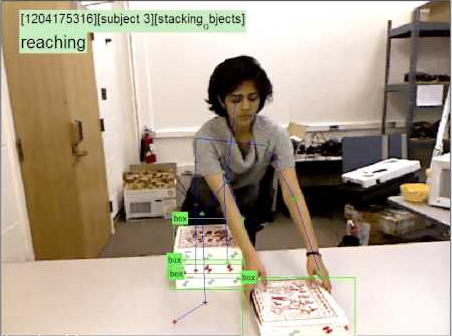}
\label{fig:cad120} }\quad
\caption{\textbf{Datasets.} a) ACTICIPATE motion and eye gaze dataset. b) CAD120 RGB-D motion dataset.}
\label{fig:datasets}
\vspace{-1em}
\end{figure}

The feature importance is evaluated on a combined gaze and skeleton dataset which was acquired and published in the ISR Vislab ACTICIPATE\footnote{The ACTICIPATE dataset can be downloaded from the following web page: http://vislab.isr.tecnico.ulisboa.pt/datasets/} project (Fig. \ref{fig:acticipate}). This dataset consists of a human actor's gaze and skeleton movement while performing either one of six actions (Place Left, Place Center, Place Right, Give Left, Give Center, Give Right). This dataset was recorded using the Optitrack motion capture system, and Pupil Labs binocular eye gaze tracking system, synchronised at a 120Hz frequency. The total number of action sequences is 120. The sequences have an average length of 220 frames. Every sequence corresponds to one action and is labelled accordingly. 

The multiple action sequence prediction model is evaluated on the CAD120 dataset (Fig. \ref{fig:cad120}, \cite{Koppula2012}). This dataset consists of a human actor's skeleton movement while performing a sequence of actions like ``pouring" and ``eating". This dataset is of special interest since it covers the scope of action sequences and it is not limited to one action per video segment. It is one of the few datasets which has a varying order of action sequences. This dataset consists of joint position and orientation feature sequences together with the respective action labels at a sample frequency of 5Hz. The total number of action sequences is 120 and the sequences have an average length of 25 time steps. 

\subsection{Feature Importance}

In our first experiment, we train the model on the combined body pose and gaze features to confirm that it yields the expected behaviour. As the movement progresses, the model receives more information and identifies the intention, correctly converging to the true label, 5. The whole movement takes 220 frames (about 2 seconds). The model is able to predict the intention target after seeing less than half of the total trajectory, about 100 frames.  
\begin{figure}[htbp]
	\centering
	\includegraphics[clip, trim=4cm 11cm 1cm 11.5cm, scale=0.7]{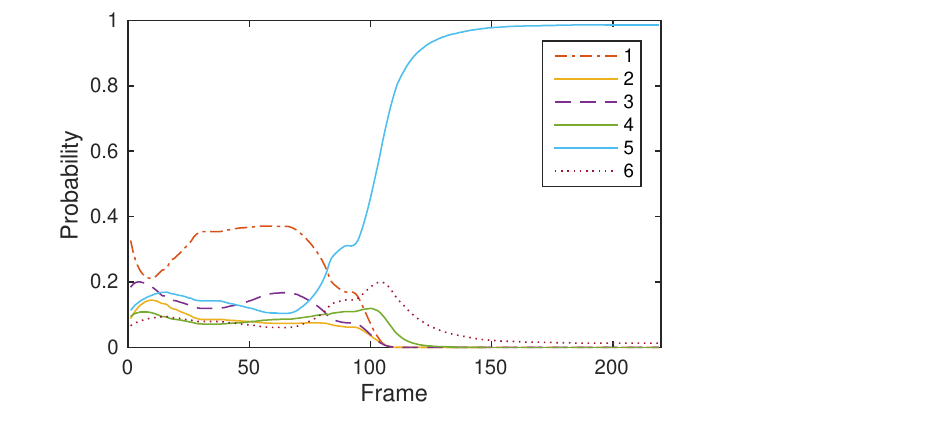}
	\caption{\textbf{Action probability temporal evolution.} The model starts with uniform probability and after about 100 frames converges to the correct label.}
	\label{distribution}
	\vspace{-1em}
\end{figure}

\begin{figure}[htbp]
	\centering
	\includegraphics[clip, trim=4.5cm 11cm 2cm 11.5cm, scale=0.70]{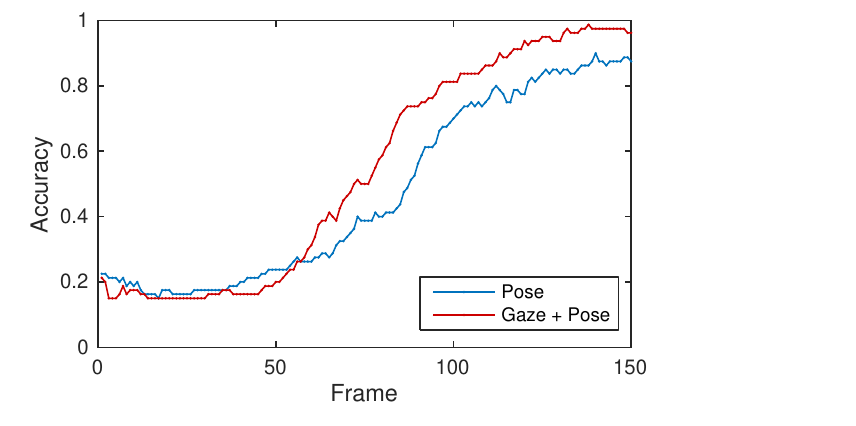}
	\caption{\textbf{Gaze and pose accuracy.} Accuracy of a model trained on (i) pose only features, and (ii) trained on combined gaze and body pose features.}
	\label{gazevsjoint}
	\vspace{-1em}
\end{figure}

The second experiment is concerned with quantifying the relative importance of the different non-verbal cues in predicting human intent. The model is trained on two sets of features: (i) combined gaze and pose cues, and (ii) body pose only.   Fig.~\ref{gazevsjoint} shows the model performance under these two conditions and the importance of the gaze information for the correct prediction of human action.  

The difference in accuracy between the two sets of cues hints at the importance of gaze. Despite the model performing similarly with and without gaze, the results show that gaze has an important role in early prediction of human activity. The model trained on both gaze and body pose cues predicts the correct action 92 ms before the model with only body cues. An interesting result is that this delay coincides with the range of delays between eye and hand movement observed in research on eye-arm movement coupling \cite{Angel1970}. 

Having established the relative importance of both gaze and body pose features in action prediction, in the next section we will evaluate the multiple action sequence prediction model on a multi-action pose feature dataset. 

\subsection{Prediction Model}

The model takes the pose features, observed over three time steps, as input in order to predict future actions as accurately as possible. 
%
% Here the model is trained on a 3 step past action history with a future reference 8 step action trajectory.
%
We will investigate how the prediction model's parameters affect the performance. The model is evaluated on the CAD120 dataset, introduced before. 

Performance will be assessed with the F1 score \cite{VanRijsbergen1979}. 
The F1-score is evaluated on a 4-fold cross validation scheme, with the final score being an average over the folds' results. As there are folds without instances of some label, the F1 score is calculated directly on the true positive, false negative and false positive rate (\ref{eq:f1}): 
\begin{equation}\label{eq:f1}
	F1 = \frac{2\cdot \text{TruePos}}{2\cdot \text{TruePos} + \text{FalseNeg} + \text{FalsePos}}.
\end{equation}

\begin{figure}[htbp]
	\centering
	\includegraphics[clip, trim=4.5cm 11cm 0cm 11.5cm, scale=0.7]{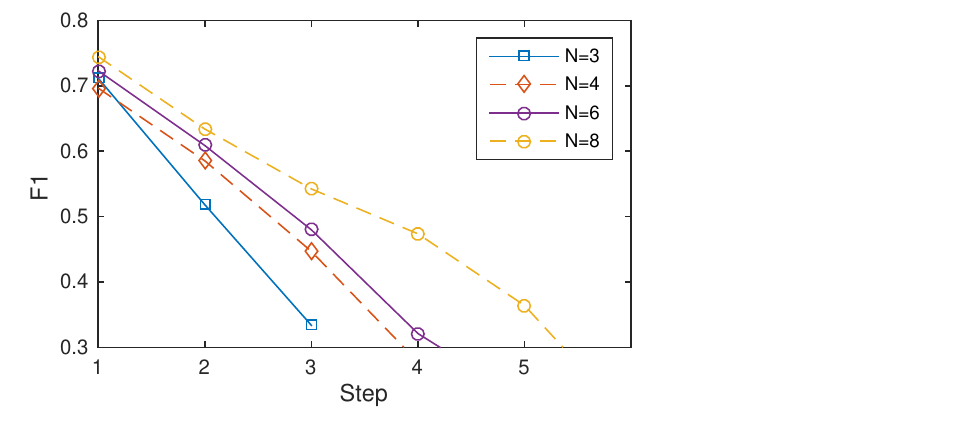}
	\caption{\textbf{Accuracy as a function of prediction length.} Prediction accuracy across time steps is positively correlated with the prediction length the model is trained on. (N corresponds to the prediction length used for training the model, Step the position in the predicted sequence.)}
	\label{predlength_acc}
	\vspace{-1em}
\end{figure}

While the model is dynamic in its ability to predict variable length action sequences, the accuracy of the action sequence prediction is influenced by the prediction length the model is trained on (Fig.~\ref{predlength_acc}). This correlation is related to the ability of the decoder to manage its internal state. When the network is trained on a long future action sequence, it learns to keep and manage the decoder's internal state, predicting longer sequences with more accuracy.

\begin{figure}[htbp]
	\centering
	\includegraphics[clip, trim=4cm 11cm 0cm 11.5cm, scale=0.65]{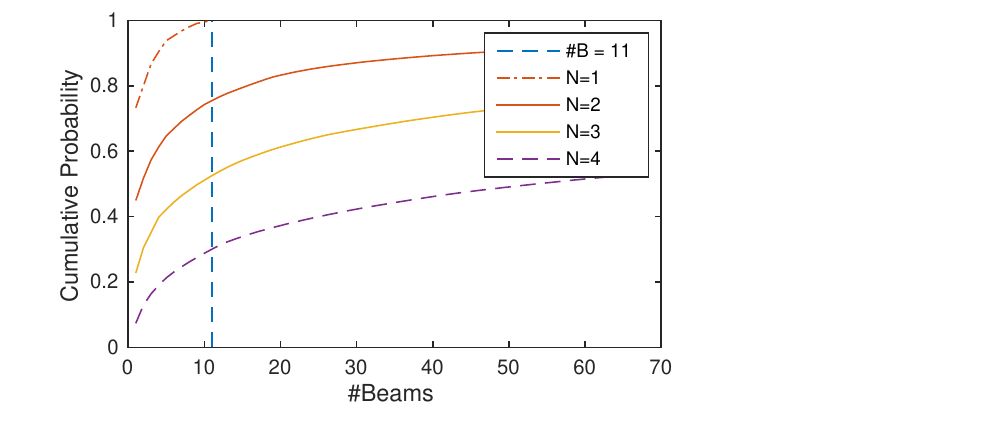}
	\caption{\textbf{Beam cumulative probability.} Cumulative probability of the outcome space the model is able to capture. "N" represents the length of the predicted trajectory, and "\#Beams" the length of the predicted action sequences.}
	\label{beam_cumsum}
	\vspace{-1em}
\end{figure}

The second parameter to analyse is the number of beams (action sequences) which determine the space of action sequences that the model is able to capture (Fig. \ref{beam_cumsum}). The cumulative sum of the beams' probabilities is a measure of the solution space that we are able to cover with a given number of beams.

The space of possible solutions grows exponentially with the number of prediction steps. While a beam width of 11 beams is able to capture 100\% of the outcome probability space in a one-step ahead prediction scenario, the same number of beams only captures around 75\% of the outcome probability space in the two-step ahead prediction scenario. As the solution space grows, a fixed number of beams captures a cumulative probability outcome space that decays with the number of prediction steps.

%Since the distributions empirically follow a long tail distribution, the cumulative probability with a fixed beam size decreases linearly with the number of prediction steps. 

%

\begin{figure}[htbp]
	\centering
	\includegraphics[clip, trim=5cm 10.5cm 5cm 10.5cm, scale=0.55]{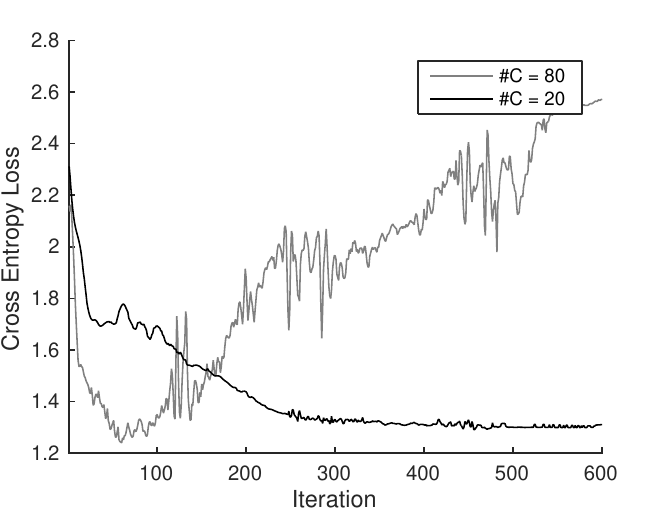}
	\caption{\textbf{Validation loss as a function of the context dimensionality.} The iteration represents the number of training steps, while \#C represents the dimensionality of the context vector parameter. As the dimensionality parameter is increased, the network starts to overfit to the training data.}
	\label{hidden_state_size}
	\vspace{-1em}
\end{figure}

It is well known that the generalization error is related to the model's capacity, the ability to learn complex patterns \cite{Goodfellow2016}. The dimensionality of the context vectors is a parameter which defines the model's capacity. Increasing this dimension reduces the informational bottleneck, increasing the model's capacity and as a consequence the generalization error. Increasing the generalization error makes the model prone to over fitting to the training set and not generalizing to new samples (Fig. \ref{hidden_state_size}). Hence, the context vector dimensionality acts as a regularizer of the model.

% The final F1 accuracy for the model trained on a 8 step prediction length is summarized in table \ref{acc_table}. 

% \begin{table}[htbp] 
% 	\def\arraystretch{2}
% 	\centering
% 	\caption{F1 Score as a function of the prediction step.}
% 	\label{acc_table}
% 	\begin{tabular}{|c|c|c|c|c|c|c|}
% 		\hline
% 		Step              & 1     & 2     & 3     & 4     & 5     & 6     \\ \hline
% 		F1  (\%) & 74.49 & 63.40 & 54.23 & 47.36 & 36.39 & 18.63 \\ \hline
% 	\end{tabular}

% \end{table}

%A metric which condenses the model's predictive ability is the confusion matrix. This matrix enables us to visually understand classification errors such as false positives and false negatives.

%\begin{figure}[htbp]
%	\centering
%	\includegraphics[clip, trim=2cm 8cm 2cm 8.5cm, scale=0.3]{confusion_matrix}
%	\caption{Action prediction confusion matrix. [Faltam legendas]}
%	\label{confusion_matrix}
%\end{figure}

\section{CONCLUSIONS}

%We addressed the problems of non-verbal gaze and body pose cues importance, variable length action sequence prediction and estimation of multiple possible future action sequences in the context of human-robot cooperation.

We showed the importance of both body pose and gaze cues for the accurate prediction of human intent. More specifically, the experiments demonstrated that a model trained on both body and gaze cues predicts the correct action about 92ms before a model trained only on body pose cues.

We introduced a recurrent neural network topology designed to predict multiple and variable length action sequences. Predicting action sequences introduces combinatorial complexity issues which were successfully mitigated using a pruning method. 

We demonstrated the theoretical value of predicting multiple and variable action sequences for estimating the expected future reward in a human robot cooperation scenario.

We studied how different training procedure and parameter combinations affect the model performance. All tests were carried out on realistic publicly available datasets.

%: the training prediction length is positively correlated with the model accuracy; the number of predicted action sequences defines the outcome space that can be captured; and the context vectors dimensionality is a parameter defining the model capacity.

Our approach extends the state of the art in directions that are key to enable more efficient human-robot cooperation, particularly involving non-verbal communication.

\addtolength{\textheight}{0cm}   % This command serves to balance the column lengths
                                  % on the last page of the document manually. It shortens
                                  % the textheight of the last page by a suitable amount.
                                  % This command does not take effect until the next page
                                  % so it should come on the page before the last. Make
                                  % sure that you do not shorten the textheight too much.

%%%%%%%%%%%%%%%%%%%%%%%%%%%%%%%%%%%%%%%%%%%%%%%%%%%%%%%%%%%%%%%%%%%%%%%%%%%%%%%%

%%%%%%%%%%%%%%%%%%%%%%%%%%%%%%%%%%%%%%%%%%%%%%%%%%%%%%%%%%%%%%%%%%%%%%%%%%%%%%%%

%%%%%%%%%%%%%%%%%%%%%%%%%%%%%%%%%%%%%%%%%%%%%%%%%%%%%%%%%%%%%%%%%%%%%%%%%%%%%%%%
%\section*{APPENDIX}

\section{FUTURE WORK}

Possible directions include extending the model by exploring the connection between non-verbal cues and semantic features related to the context, through composing the model with additional information using probabilistic methods.

Furthermore, this work establishes a strong base for the implementation of a joint action scenario on a humanoid robotics platform such as the iCub. 

\section*{ACKNOWLEDGMENTS}

Research supported by the Portuguese Foundation for Science and Technology (FCT) project [UID/EEA/50009/2013], EU H2020 project under Grant 752611 - ACTICIPATE and RBCog-Lab research infrastructure.

%%%%%%%%%%%%%%%%%%%%%%%%%%%%%%%%%%%%%%%%%%%%%%%%%%%%%%%%%%%%%%%%%%%%%%%%%%%%%%%%

\bibliography{ICRAPaper} 
\bibliographystyle{ieeetr}

\end{document}